\documentclass[10pt, a4paper, twocolumn]{article}

\title{Black holes in Einstein-Gauss-Bonnet gravity with a string cloud background}
\author{Estanislao Herscovich\footnote{Institut f\"ur Mathematik, Universit\"at Paderborn, 33098,
Paderborn, Germany. E-mail: \href{mailto:herscovi@math.uni-paderborn.de}{herscovi@math.uni-paderborn.de}}
 ~and Mart\'{\i}n G. Richarte\footnote{Departamento de F\'{\i}sica, Facultad de Ciencias Exactas y Naturales, UBA Ciudad Universitaria,
Pabell\'on I (1428), Buenos Aires, Argentina. E-mail: \href{mailto:martin@df.uba.ar}{martin@df.uba.ar}}} 
\date{\today}

\input{xy}
\xyoption{all}
\xyoption{poly}
\usepackage[all]{xy}
\usepackage{amsmath,amsthm,amsfonts,amssymb,stmaryrd,fancyhdr}
\usepackage{abstract}
\usepackage{latexsym}
\usepackage{color}
\usepackage[dvips]{graphicx}
\usepackage{float}  
\usepackage{fullpage}
\usepackage[T1]{fontenc}
\usepackage{palatino}

\usepackage[
  breaklinks,
  naturalnames,
  ]{hyperref}

\newtheorem{teo}{Theorem}[section]

\newtheorem{lema}[teo]{Lemma}

\numberwithin{equation}{section}                    

\newcommand{\ben}{\begin{eqnarray}}
\newcommand{\een}{\end{eqnarray}}

\newcommand\RR{{\mathbb{R}}}

\newcommand\ZZ{{\mathbb{Z}}}
\newcommand\NN{{\mathbb{N}}}

\def\L{{\mathcal L}}

\begin{document}

\twocolumn[
\maketitle 
\begin{onecolabstract} 
We obtain a black hole solution in the Einstein-Gauss-Bonnet theory for the string cloud model in a five dimensional spacetime. 
We analyze the event horizons and naked singularities. 
Later, we compute the Hawking temperature $T_{\mathrm{H}}$, the specific heat $C$, the entropy $S$, and the Helmholtz free energy $F$ of the black hole. 
The entropy was computed using the Wald formulation. 
In addition, the quantum correction to the Wald's entropy is considered for the string cloud source. 

We mainly explore the thermodynamical global and local stability of the system with vanishing or non-vanishing cosmological constant. 
The global thermodynamic phase structure indicates that the Hawking-Page transition is achieved for this model. 
Further, we observe that there exist stable black holes with small radii and that these regions are enlarged when choosing small values of the string cloud density and of the Gauss-Bonnet parameter. 
Besides, the rate of evaporation for these black holes are studied, determining whether the evaporation time is finite or not. 
Then, we concentrate on the dynamical stability of the system, studying the effective potential for s-waves propagating on the string cloud background. 
\end{onecolabstract}
]
\saythanks 

\section{Introduction}

Black holes are one of the most exciting and fascinating objects on gravity physics, 
mostly because their thermodynamical properties have a deep similarity with statistical mechanics \cite{Ha1}, \cite{Be1}, \cite{BCH1}.
In the last decades there has been a renewal of interest in higher dimensional spacetimes associated with this kind of objects. 
Moreover, a lot of effort to understand the quadratic corrections to the Einstein-Hilbert action has been put. 

At the present time, it is well-known that for a spacetime with dimension $D \geq 5$ the Einstein-Hilbert action of gravity admits quadratic corrections constructed from coordinate-invariant tensors which scale as fourth derivatives of the metric. 
In particular, when $D = 5$, the most general theory leading to second order equations for the metric is the so-called \emph{Einstein-Gauss-Bonnet (EGB) theory} or \emph{Lovelock theory} up to second order. 
This class of model for higher dimensional gravity has been widely studied, because it can be obtained in the low energy limit of string theory 
(see \cite{Z1}). 
For a $D$-dimensional spacetime with $D < 5$, the \emph{Gauss-Bonnet (GB)} term does not give any contribution to the dynamical equations.

According to the recent theoretical developments, a scenario in which the fundamental building blocks of nature are extended objects instead of point objects should be considered quite seriously. 
In particular, one-dimensional objects (strings) are the most popular candidates. 
The study of the gravitational effects of matter in the form of clouds of both cosmic and fundamental strings has then deserved considerable attention 
(see \cite{BY1}, \cite{Le1}, \cite{So1}). 
Many authors have found exact black hole solutions in Einstein-Gauss-Bonnet models with different kind of sources, for instance, with Maxwell and Born-Infeld energy-momentum tensor, matter, a radiating source and Yang-Mills fields 
(see \cite{DG1}, \cite{DOT1}, \cite{MD1}, \cite{Wi1}, \cite{MHM1}, \cite{HMH1}). Moreover, black holes  with a five dimensional  Gauss-Bonnet
term  in the bulk and an induced three-dimensional
curvature term on the brane were also found \cite{CPTZ1}, \cite{CPTZ2}.

Also, some authors have put great attention to the thermodynamics of these solutions, as a way to characterize the black holes  (see \cite{CEJM2}, \cite{CN1}, \cite{CNO1}, \cite{CG1}, \cite{DMMS1}, \cite{NO1}). 
In particular, R. Cai has studied the phase structure of the topological black holes in the EGB theory with negative cosmological constant 
(see \cite{Cai1}). 
More recently, in \cite{CRS1} the authors have investigated the entropy of black holes in GB and Lovelock gravity using the Noether charge approach. 
In addition, in \cite{MKP1} the authors have explored all the thermodynamics quantities for the EGB model with AdS backgrounds. 
They have concentrated on the role of the sign of the Gauss-Bonnet coupling constant $\alpha$. 

We are interested in the construction of black hole solutions in a five dimensional spacetime with an energy-momentum tensor coming from a 
string cloud model. 
In the present work our starting point is to solve the full non-linear Einstein-Gauss-Bonnet equations for a static and spherically symmetric metric. 
We study not only the structure of the geometry but also the thermodynamical stability of the system. 
Later, we shall concentrate on the dynamical stability, \textit{i.e.} we will study the effective potential arising from the scalar waves perturbation satisfying either the massive or non-massive Klein-Gordon equation. 

The paper is organized as follows. 
In Section \ref{sec:gen} we recall the basic facts of the Einstein-Gauss-Bonnet theory and of the string cloud model 
studied by P. Letelier. 
In the next section we present the exact solution for a EGB theory on a static spherically symmetric $5$-dimensional spacetime with a source provided by a string of clouds. 
As a consistency check, we reobtain the well-known Schwarzschild $5$-dimensional spacetime from our solution 
as a particular case. 
Moreover, we also study the possible event horizons and singularities.  In Section \ref{sec:termo} we focus on the thermodynamical properties of the aforementioned spacetime. 
More concretely, we obtain the Hawking temperature, the Helmholtz free energy, the heat capacity and the entropy. In particular, the entropy of the black hole was computed using the Wald prescription.
The sign of the heat capacity  allows us to determine the local stability of the system, whereas the sign of the free energy 
tells us if the system is globally preferred or not. 

In Subsection \ref{subsec:quantum}, following th techniques of the tunneling method considered in \cite{BM1}, we compute the quantum 
corrections to the entropy, and we prove that it coincides with the standard expressions (\textit{cf.} eq. (27) in \cite{ZRL1}) 
under fairly more general assumptions. 
In the next subsection we explicitly determine when a black hole evaporates in a finite time, 
generalizing previous considerations in the literature.

Finally, in Section \ref{sec:scalar} we consider a scalar field propagating in the bulk given by 
the spacetime and we determine the effective potential in this case and its third order approximation.


\section{Generalities}
\label{sec:gen}

\subsection{The Einstein-Gauss-Bonnet theory}

We first recall the general facts about the Einstein-Gauss-Bonnet model on five dimensions. 
It is a generalization of Einstein Relativity, since its action also involves quadratic corrections constructed from coordinate-invariant tensors 
which scale as fourth derivatives of the metric (see \cite{Lov1}). 
The action for this theory coupled on a $5$-di\-men\-sion\-al Lorentzian manifold $(M,g)$ with matter fields is as follows 
\[     S = \frac{1}{2}\int_{M} \sqrt{-\mathbf{g}} \big(R - 2 \Lambda + \alpha R^{2}_{\mathrm{GB}}\big) d^{5}x 
         + S_{\mathrm{mat}},      \]
where $S_{\mathrm{mat}}$ denotes the action associated with the matter fields, $\alpha$ is the so-called \emph{Gauss-Bonnet coupling constant}, which we consider to be non-negative, 
$\mathbf{g}$ is the determinant of the metric $g$ and 
\[     R^{2}_{\mathrm{GB}} = R_{\mu \nu \rho \sigma} R^{\mu \nu \rho \sigma} + R^{2} - 4 R_{\mu \nu} R^{\mu \nu}.     \] 
Hereafter  we will adopt the signature $(-,+,+,+,+)$ for the metric and we shall follow all the main conventions of General Relativity (GR) explained in \cite{SW1}. 

The corresponding Euler-Lagrange equations (also called \emph{EGB equations}) for the metric are 
\begin{equation}
\label{eq:eqegb}
     \mathcal{G}_{\mu \nu} = \mathcal{G}_{\mu \nu}^{0} + \mathcal{G}_{\mu \nu}^{1} + \mathcal{G}_{\mu \nu}^{2} = T_{\mu \nu},     
\end{equation}
where $T_{\mu \nu}$ is the energy-momentum tensor and 
\begin{align*}
   \mathcal{G}_{\mu \nu}^{0} &= \Lambda g_{\mu \nu}, \hskip 5mm   \mathcal{G}_{\mu \nu}^{1} = R_{\mu \nu} - \frac{1}{2} R g_{\mu \nu},
   \\
   \mathcal{G}_{\mu \nu}^{2} &= - \alpha \Big( \frac{1}{2} g_{\mu \nu} R^{2}_{\mathrm{GB}} - 2 R R_{\mu \nu} + 4 R_{\mu \rho} 
   R^{\rho}_{\phantom{\rho} \nu} 
   \\ 
   &\phantom{= - \alpha} + 4 R_{\rho \sigma} R^{\rho \sigma}_{\phantom{\rho \sigma} \mu \nu} - 2 R_{\mu \rho \sigma \eta} 
   R^{\phantom{\nu} \rho \sigma \eta}_{\nu} \Big).
\end{align*}
We point out that, as in the plain General Relativity, the divergence of the EGB tensor $\mathcal{G}_{\mu \nu}$ vanishes. 

As a remarkable property of this theory we may say that the resulting equations of motion have no more than second order derivatives 
of the metric and the theory has been shown to be free of ghosts when it is expanded around the flat space, 
avoiding any problem concerning unitarity (see \cite{BD1}). 
Further, it has been argued that the Gauss-Bonnet term appears as the leading correction to the effective low-energy action 
of the heterotic string theory (see \cite{Z1}, \cite{Ne1}).

\subsection{The string cloud model}

In this subsection we shall briefly review the theory of clouds of strings (see \cite{Le1}). 
This cloud of strings will be the source of the subsequent families of spacetimes satisfying the Einstein-Gauss-Bonnet model.

The Nambu-Goto action of a string evolving in the base manifold $(M,g)$ is 
\[    S_{\mathrm{NG}} = \int_{\Sigma} p \sqrt{-\mathbf{h}} d\lambda^{0} d\lambda^{1},     \]
where $p$ is a positive constant related to the string tension, 
$(\lambda^{0}, \lambda^{1})$ is a parametrization of the world sheet $\Sigma$ described by the string with induced metric given by 
\[     h_{a b} = g_{\mu \nu} \frac{\partial x^{\mu}}{\partial \lambda^{a}} \frac{\partial x^{\nu}}{\partial \lambda^{b}}     \]
and $\mathbf{h}$ is the determinant of $h$.

Alternatively, we may describe the world sheet by a bivector $\Sigma \in \Gamma(\Lambda^{2}TM)$ of the form 
\begin{equation}
\label{eq:bivector}
     \Sigma^{\mu \nu} = \epsilon^{a b} \frac{\partial x^{\mu}}{\partial \lambda^{a}} \frac{\partial x^{\nu}}{\partial \lambda^{b}},     
\end{equation}
where $\epsilon^{a b}$ denotes the two-dimensional Levi-Civita tensor given by $\epsilon^{0 1} = - \epsilon^{1 0} = 1$. 
The former satisfies the following identities 
\begin{equation}
\label{eq:imp}
   \Sigma^{\mu [\alpha} \Sigma^{\beta \gamma]} = 0, \hskip 5mm \nabla_{\mu} \Sigma^{\mu [\alpha} \Sigma^{\beta \gamma]} = 0,     
\end{equation}
where the square brackets indicate antisymmetrization in the enclosed indices. 
By the Frobenius' theorem, any bivector satisfying the previous identities leads to a parametrized surface such that \eqref{eq:bivector} holds. 

We would like to point out the following identity 
\begin{equation}
\label{eq:usable}
     \Sigma^{\mu \sigma} \Sigma_{\sigma \tau} \Sigma^{\tau \nu} = \mathbf{h} \Sigma^{\nu \mu},
\end{equation}
which will be used in the sequel and which follows directly from the definition of $\mathbf{h}$ and $\Sigma$. 

In this description, the Lagrangian density becomes 
\[     \L = p \Big(-\frac{1}{2} \Sigma^{\mu \nu} \Sigma_{\mu \nu}\Big)^{1/2}.     \]
Taking into account that $T^{\mu \nu} = 2 \partial \L/\partial g^{\mu \nu}$, we may obtain that the energy-momentum tensor for one string is 
given by $T^{\mu \nu} = p \Sigma^{\mu \rho} \Sigma_{\rho}^{\phantom{\rho} \nu}/\sqrt{-\mathbf{h}}$. 
Hence, we shall consider the following energy-momentum tensor 
\[     T^{\mu \nu} = \rho \frac{\Sigma^{\mu \sigma} \Sigma_{\sigma}^{\phantom{\sigma} \nu}}{\sqrt{-\mathbf{h}}}     \]
for a string cloud with density $\rho$. 

By making use of the Bianchi identity and the fact that the divergence of the Einstein tensor vanishes we see that 
$(-\mathbf{h})^{-1}\nabla_{\mu} (\rho \Sigma^{\mu \sigma}) \Sigma_{\sigma}^{\phantom{\sigma} \nu} \Sigma_{\nu \tau} = 0$. 
Contracting the previous identity with $\Sigma_{\tau \nu}$ and using equation \eqref{eq:usable}, we obtain 
$\nabla_{\mu} (\rho \Sigma^{\mu \sigma}) \Sigma_{\sigma}^{\phantom{\sigma} \nu} = 0$. 
Employing a system of coordinates adapted to the parametrization of the surface, we see that in fact 
\begin{equation}
\label{eq:divtens2}
     \partial_{\mu} (\sqrt{-\mathbf{g}} \rho \Sigma^{\mu \sigma}) = 0.     
\end{equation}

\section{Exact solutions}
\label{sec:exacsol}

From now on, we shall restrict ourselves to the case of static spherically symmetric $5$-dimensional EGB spacetimes $(M,g)$, so we may suppose that (locally) there exists a system of coordinates $(t,r,\phi,\theta,\xi)$ such that the metric $g$ is given by
\[     g_{\mu \nu} dx^{\mu} \otimes dx^{\nu} 
       = - G(r) dt \otimes dt + \frac{1}{G(r)} dr \otimes dr + r^{2} \omega,     \]
where $G$ is a real-valued non-vanishing $C^{\infty}$ function of $r$ and $\omega$ is the volume form of $S^{3}$ for the remaining coordinates, namely, 
\[     \omega = d\xi \otimes d\xi + \sin^{2}(\xi) (d\theta \otimes d\theta + \sin^{2}(\theta) d\phi \otimes d\phi).      \]

We shall also assume that the string cloud is spherically symmetric, so the only non-vanishing component may be $\Sigma^{t r} = - \Sigma^{r t}$, 
because $\Sigma$ is a bivector. 
Therefore, the  non-zero components of bivector are given by
\[     \Sigma^{\nu\mu} = \Sigma^{0}_{\phantom{0}1}(r)\Big(\delta^{\nu}_{\phantom{\nu}0}\delta^{\mu}_{\phantom{\mu}1}-\delta^{\mu}_{\phantom{\mu}0}\delta^{\nu}_{\phantom{\nu}1}\Big).      \]

This source corresponds physically to a collection of strings which are extended along the radial direction, 
distributed uniformly over the sphere. 
That is, the worldsheet of each string covers the $t$-$r$ plane, which is the simplest case. 

\subsection{\texorpdfstring{Computation of $T_{\mu \nu}$}{Computation of the energy momentum tensor}}

Since the only possible non-vanishing component of the bivector $\Sigma$ is $\Sigma^{t r} = - \Sigma^{r t}$, we have
\[     T^{t t} = - \frac{\rho}{\sqrt{-\mathbf{h}}} \Sigma^{t r} \Sigma_{r}^{\phantom{r} t}.     \] 
Taking into account that $\mathbf{h} = \frac{1}{2} 2 \Sigma^{t r} \Sigma_{t r} = - (\Sigma^{t r})^{2}$, 
we conclude that 
$T^{t}_{\phantom{t} t} = - \rho |\Sigma^{t r}|$, $T^{r}_{\phantom{r} r} = - \rho |\Sigma^{t r}|$, 
and the other components vanish. 
Furthermore, the identity \eqref{eq:divtens2} yields $\partial_{r}(r^{3} T^{t}_{\phantom{t} t}) = 0$. 
As a consequence, 
\begin{equation}
\label{eq:tensfinal}
   T^{r}_{\phantom{r} r} = T^{t}_{\phantom{t} t} = - \frac{a}{r^{3}},
   \footnote{In our convention, we easily see that that the value of the tensor-energy momentum at a timelike vector 
   (\textit{e.g.} $T_{t t}$ outside the black hole or $T_{r r}$ inside it) is always positive (\textit{cf.}~\cite{SW1})}
\end{equation}
for some real constant $a$, that we will take to be non-negative for physical reasons.

We shall now discuss the energy conditions for this source. 
For the the energy-momentum tensor written in diagonal form as $T^{\mu}_{\phantom{\mu} \nu} = \mathrm{diag}(-\sigma, p_{r}, 0,0,0)$, 
the \emph{weak energy condition (WEC)} means that $\sigma \geq 0$, and $p_{r} + \sigma \geq 0$, 
whereas the \emph{dominant energy condition (DEC)} says that $\sigma \geq 0$, and $|p_{r}| \leq \sigma$. 
Moreover, the \emph{strong energy condition (SEC)} states that $\sigma + p_{r} \geq 0$ and $3 p_{r} + \sigma \geq 0$ (\textit{cf.}~\cite{Po1}). 
The physical meaning of $\sigma$ and $p_{r}$ is the energy density and the radial pressure, respectively. 
It is important to mention that the WEC implies the so called \emph{null energy condition (NEC)} which says that $\sigma + p_{r} \geq 0$ (\textit{cf.}~\cite{Po1}). 

In our model we obtain that $\sigma = a/r^3$ and $p_{r} = -a/r^{3}$. 
We see thus that the NEC is always satisfied.
Also, the WEC and the DEC are satisfied for $a \geq 0$, otherwise these conditions are violated. 
On the other hand, we find that the SEC is violated for $a > 0$, 
whereas for the opposite case the SEC holds. 

\subsection{\texorpdfstring{Computation of the metric $g_{a b}$}{Computation of the metric}}
\label{subsec:g}

The diagonal EGB equations \eqref{eq:eqegb} for the energy momentum tensor given in the previous subsection are 
\begin{equation}
\label{eq:lovelock}
   \mathcal{G}^{t}_{\phantom{t} t} = \mathcal{G}^{r}_{\phantom{r} r}  
   = - \frac{a}{r^{3}}, 
   \hskip 5mm
   \mathcal{G}^{\phi}_{\phantom{\phi} \phi} = 
   \mathcal{G}^{\theta}_{\phantom{\theta} \theta} 
   = \mathcal{G}^{\xi}_{\phantom{\xi} \xi}
   = 0.
\end{equation}
After a rather long manipulation we obtain that the components of the EGB tensor are given by 
\begin{align*}
 \mathcal{G}^{t}_{\phantom{t} t} &= \frac{1}{2 r^{3}} \Big(2 r^{3} \Lambda + 6 r G - 6 r + 3 r^{2} G'\Big)
\\
&+ \frac{6}{ r^{3}}\alpha G' (1 - G),
\\
 \mathcal{G}^{\phi}_{\phantom{\phi} \phi} &= -1 + r^{2} \big(\Lambda + \frac{G''}{2}\big) + G +2rG'-2 \alpha G'^{2}
 \\
 &+ 2 \alpha G'' (1 - G),
\end{align*}
and the remaining EGB equations are trivial. 
We point out that the second family of differential equations \eqref{eq:lovelock} 
are a consequence of the first family in \eqref{eq:lovelock}.  
In consequence, $G(r)$ satisfies the identity 
\[     G^{2} - 2 r^{2} G + \frac{1}{6\alpha}\Big[ 3 r^{2} - 2 a r + b - \frac{r^{4}\Lambda}{2}\Big] = 0,     \]
for some real constant $b$. Finally, there are two branches for the solution 
\begin{equation}
\label{eq:dossol}
     G(r) = 1 + \frac{r^{2}}{4 \alpha} 
     \Big(1 \pm \sqrt{\bar{\Delta}}\Big),     
\end{equation}
where
\[     \bar{\Delta} = 1+\frac{8\alpha (2\alpha-3 b)}{9 r^{4}} + \frac{4}{3} \alpha \Lambda 
       + \frac{16 a \alpha}{3 r^{3}}.     \]
Taking into account that a realistic physical solution must become the classical Schwarzschild solution of General Relativity 
in five dimensions without sources (\textit{i.e.} for $\Lambda = 0$ and $a = 0$) when considering the limit $\alpha \rightarrow 0$, 
we conclude that the solution \eqref{eq:dossol} with plus sign has no physical interest, since in that case the mentioned limit does not 
exist\footnote{A detailed analysis of the positive branch with $a=0$, and $\Lambda=0$ was made in~\cite{MS1}}. 
Moreover, from the comparison of the expansion up to first order terms of the square root of the solution \eqref{eq:dossol} 
and the classical Schwarzschild solution of general relativity (with $a=0$ and $\Lambda = 0$) we conclude that 
\begin{equation}
\label{eq:solfinal}
     G(r) = 1 + \frac{r^{2}}{4 \alpha} \left(1 - \sqrt{1+\frac{16\alpha m}{r^{4}} + \frac{4}{3} \alpha \Lambda 
       + \frac{16 a \alpha}{3 r^{3}}} \right).     
\end{equation}

In general, the solution may have two possible singularities: the usual singularity at $r=0$ and also the so-called \emph{branch singularity} at $r_{b}>0$ which is defined as follows. 
Since $G(r)$ must be real-valued, we should only consider the domain values $r$ such that the radicand in \eqref{eq:solfinal} is non-positive. 
If $1 + 4 \alpha \Lambda / 3 \geq 0$ this radicand is always non-negative, but for $1 + 4 \alpha \Lambda / 3 < 0$ there exist 
values of $r$ such that it is negative.
If  $1 + 4 \alpha \Lambda / 3 < 0$ holds, we define $r_{b}$ as the minimum positive real root of the quartic polynomial 
\begin{equation}
\label{eq:bs}
   r^{4}_{b}\big(1+\frac{4}{3}\alpha\Lambda\big)+ 16\alpha\big( \bar{a}r_{b}+  m\big)=0,  
\end{equation}
if it has three different positive roots, or as the maximum positive root, otherwise.
Hence, the metric is a priori only defined for $r \in (0,r_{b})$. 
So, as we said, if $1 + 4 \alpha \Lambda / 3 < 0$, there is a finite radius a singularity and the allowed domain of the radial coordinate is then no more $0 < r < \infty$ (\textit{cf.}~\cite{TM1}). 

On the other hand, if $r \rightarrow 0$, $G(r)$ approaches to $1-\sqrt{m/\alpha}$, which can be interpreted as the fact that 
the EGB term removes the metric singularities at the origin (\textit{cf.}~\cite{Wi1}). 

In a similar way, for the vacumm state (\textit{i.e.} $m=0$, $a=0$) for \eqref{eq:dossol} the metric takes the form 
\[     G(r) = 1+\frac{r^{2}} {2\alpha}\Big(1 - \sqrt{1+\frac{4}{3}\alpha\Lambda }~\Big).     \]
In order that $G(r)$ is real-valued, we must have that $\alpha\Lambda\geq-3/4$. 
Hence, if $\Lambda \geq 0$ this condition is fulfilled since $\alpha \geq 0$, whereas for a negative  cosmological constant $\Lambda$, the GB coupling constant $\alpha$ must be in the interval $ [0, -3/(4\Lambda)]$.

If $1 + 4 \alpha \Lambda / 3 <  0$, then the expansion of $G(r)$ near $r_{b}$ is 
\[     G(r) \simeq \Big( 1 + \frac{r_{b}^{2}}{4 \alpha} \Big) - \frac{1}{\alpha} \sqrt{\alpha \Big( \frac{4 m}{r_{b}} + a \Big)} (r_{b}-r)^{1/2}.     \]

Let us now suppose that $1 + 4 \alpha \Lambda / 3 \geq  0$.
We may analyze the asymptotic behaviour of the previous expression in the limit 
$r \rightarrow +\infty$, which is of the form 
\[     G(r) \simeq 1 - \frac{\Lambda}{3(1+\sqrt{1+\frac{4 \alpha \Lambda}{3}})} r^{2}.     \]   
This leads to the well-known fact that the Gauss-Bonnet term provides a correction to 
the original cosmological constant $\Lambda$ (\textit{cf.}~\cite{BD1}). 

\subsection{\texorpdfstring{Event horizons of the black hole}{Event horizons of the black hole}}

In what follows  we are going to briefly discuss the general properties of the static solutions found in the last section.
We have there studied the conditions for the function $G(r)$ to be real-valued, but in order that it defines a metric it must be non-zero. 

By definition an event horizon is a hypersurface of the form $r = r_{\mathrm{h}}$ in $(M,g)$ where $r_{\mathrm{h}}$ is determined as the maximum positive root of $G(r_{\mathrm{h}})=0$. 
Therefore, using \eqref{eq:solfinal}, $r_{\mathrm{h}}$ is given by 
\begin{equation}
\label{eq:ecparahorizonte}
     \frac{\Lambda}{12} r_{\mathrm{h}}^{4} - \frac{1}{2} r_{\mathrm{h}}^{2} + \frac{a}{3} r_{\mathrm{h}} + (m - \alpha) = 0.     
\end{equation}

From the considerations of the previous subsection, in the case $1 + 4 \alpha \Lambda / 3 < 0$, 
there may be another constraint on the existence of an event horizon, indicating that the event horizon should also satisfy the inequality $r_{h} < r_{b}$. 
We shall assume that this is always the case if we suppose that an event horizon exists and the inequality $1 + 4 \alpha \Lambda / 3 < 0$ holds. 

If $\Lambda = 0$, then equation \eqref{eq:ecparahorizonte} becomes quadratic, and its solutions are 
\begin{equation}
\label{eq:rh}
   r_{\mathrm{h}} = \bar{a} \pm \sqrt{\bar{a}^{2}+2(m-\alpha)},  
\end{equation}
where $\bar{a} = a/3$. 
In this case the black hole solution has an inner horizon (smaller positive root) and an event horizon (largest positive root or outer horizon). 
This solution exhibits a similar causal structure of charged \emph{Reissner-Nordstr\"om (RN)} black hole geometry in 
$5$ dimensions. 
As a consequence of eq. \eqref{eq:rh}, the horizons for $\Lambda = 0$ are well-defined if and only if 
\[     \bar{a}^{2} + 2 (m - \alpha) \geq 0.     \]
The special case where the  equality holds is referred to as an extreme black hole with the event horizon given by $r_{\mathrm{ext}}=\bar{a}$. 
If  $\bar{a}^{2} + 2 (m - \alpha) < 0$, the string cloud solution describe a naked singularity at $r=0$. 

Now, we establish what the leading terms of the most relevant quadratic invariants 
near the physical singularity ($r=0$) are:
\begin{align*}
   R_{a b c d}R^{a b c d} &\simeq \frac{12 m}{\alpha r^{4}} + \frac{4 a}{\alpha r^{3}} + O(r^{-2}),   
   \\
   R_{a b}R^{a b} &\simeq \frac{12 m}{\alpha r^{4}} + \frac{6 a}{\alpha r^{3}} + O(r^{-2}),     
\end{align*}

It turns out that the mass term dominates over the string cloud contribution near the singularity. 
However, the last term is smoother than the one for the $5$-dimensional Schwarzschild metric with negative mass, 
that possesses a timelike naked singularity at the origin with \emph{Kretschmann scalar} $R_{a b c d}R^{a b c d}$ 
diverging as $r^{-8}$ (\textit{cf.}~\cite{DOT1}). 
On the other hand, the Kretschmann scalar for the charged Gauss-Bonnet black hole has a leading term of the form $r^{-6}$
but it also has similar terms as in our case. 


\section{\texorpdfstring{Thermodynamics of the black hole}{Thermodynamics of the black hole}}
\label{sec:termo}

In this section we shall discuss and reckon the main thermodynamical properties of the string black holes solution within the EGB framework.
For the entire section $\bar{a}$ will be treated as an external parameter in relation with the thermodynamical considerations. 

Using the standard definition of surface gravity $\mathcal{K} = 1/2|G'(r_{\mathrm{h}})|$ for a spherically symmetric static metric 
with a static Killing horizon, it follows that 
\begin {equation}
\label{sg}
 \mathcal{K} = \frac{|r_{\mathrm{h}} - \bar{a} - \frac{\Lambda}{3} r_{\mathrm{h}}^{3}|}{r^{2}_{\mathrm{h}} + 4 \alpha}.
\end{equation}
Using the identity $T_{\mathrm{H}} = \mathcal{K}/2\pi$, we see that 
\begin{equation}
\label{eq:th}
     T_{\mathrm{H}} = \frac{|r_{\mathrm{h}} - \bar{a} - \frac{\Lambda}{3} r_{\mathrm{h}}^{3}|}{2 \pi(r^{2}_{\mathrm{h}} + 4 \alpha)}.     
\end{equation}
Since it will useful later, we remark that in the formula of the Hawking temperature we have omitted the Planck constant $\hbar$ 
in the numerator by our choice of units. 

\begin{figure}[H]
\begin{center}
\includegraphics[width=0.4\textwidth]{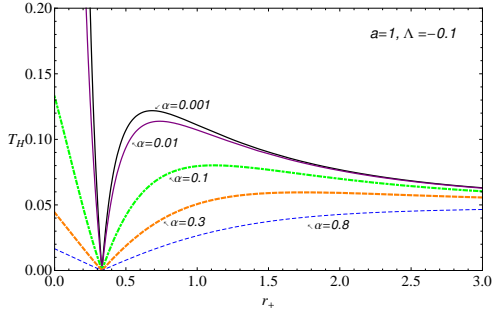}
\caption{The Hawking temperature as a function of the horizon radius for $\Lambda = -0.1$, $a = 1$ and different values of $\alpha$.} 
\label{fig:fig2bis}
\end{center}
\end{figure}

Let us consider the case with vanishing cosmological constant. 
In the limit $\alpha \rightarrow 0$ and $a \rightarrow 0$, we recover the classical result for Schwarzschild spacetime saying 
that the Hawking temperature diverges when the event horizon shrinks to zero. 
On the contrary, for vanishing $a$ but non-zero EGB coupling constant $\alpha$, 
the Hawking temperature remains finite when performing the limit $r_{\mathrm{h}} \rightarrow 0$ (\textit{cf.}~\cite{Cai1}). 
Besides, for $a \neq 0$ and $\alpha \neq 0$, there is a particular radius of the event horizon for which the Hawking temperature vanishes, 
so the black hole does not radiate energy. 
As a general behaviour, the Hawking temperature exhibits a peak, which decreases and moves to the right when the string cloud parameter $a$ grows 
(see Fig. \ref{fig:fig2bis}). 

\subsection{\texorpdfstring{Global stability}{Global stability}}
\label{subsec:global}

It is a well-known fact that asymptotically AdS Schwarzschild black holes are thermally favored when their temperature is sufficiently high, 
whereas the pure AdS background is thermally preferred for low temperatures. 

As the temperature decreases, there is a first order transition such that the black hole spacetime leads to a 
pure AdS geometry. 
This effect is known as the \emph{Hawking-Page transition (HPt)} (see \cite{HP1}). 
The phase transition has been widely studied for others asymptotically AdS black holes in the context of EGB gravity (\textit{cf.}~\cite{Cai1}, \cite{CN1}, \cite{CNO1}, \cite{CG1}, \cite{DMMS1}, \cite{NO1}). 

Now, we are going to study the thermodynamic phase structure for EGB gravity with a source given by a string cloud. 
In order to do so, we first need to express the black hole mass parameter $m$ as a function of the event horizon radius $r_{\mathrm{h}}$, which can be done using \eqref{eq:ecparahorizonte}. 
The mass parameter $m$ is related to the ADM mass $M$ of the black hole, which takes the form $M = 3 \Omega_{3} m/(8 \pi)$ for a $5$-dimensional spacetime, where $\Omega_{3}$ is the volume of the unit sphere in $\RR^{3}$. 

Our next step is to compute the entropy in the case $\Lambda \leq 0$. 
So, from the first law of thermodynamics $dM = T dS$ and the fact that $\Lambda \leq 0$, we can obtain the entropy of the black hole 
\[     S = \int_{0}^{r_{+}} \frac{dM}{dr_{+}} \frac{dr_{+}}{T(r_{+})},     \]
where $r_{+}$ denotes the radius of the largest horizon event (see \cite{CJS1}). 
We have imposed the physical assumption that the entropy vanishes when the horizon of the black hole shrinks to zero as explained by Cai 
(see \cite{Cai1}). 
The expression is given by (\textit{cf.}~\cite{MS1})
\begin{equation}
\label{eq:s}
     S = \pi \Big( \frac{r_{+}^{3}}{3} + 4 \alpha r_{+} \Big).     
\end{equation}
As it is well-known, the entropy for the EGB model does not satisfy the Bekenstein-Hawking law $S = A/4$, 
since $S = A(1 + 12 \alpha/r_{+}^{2})/4$, 
where $A = 4 \pi r_{+}^{3}/3$ is the area of the (spherical) event horizon of radius $r_{+}$ (\textit{cf.}~\cite{Cai1}). 
Interestingly, the energy-momentum tensor of the string cloud does not lead to any correction in the entropy \eqref{eq:s}. 

In the previous computation we have asssumed that the ADM mass $M$ is the energy of the black hole. 
In order to prove so, we might proceed as follows: first, compute the entropy function $S_{\mathrm{W}}$ using the Wald formulation and 
then, from the first principle of thermodynamics $dS_{\mathrm{W}} = M_{\mathrm{W}} dT_{\mathrm{H}}$, deduce the (real) mass of the black hole $M_{\mathrm{W}}$. 
We recall that the Wald formulation gives an expression of the entropy for any $N$-dimensional spacetime with a diffeomorphism invariant Lagrangian such that it admits stationary black hole
solutions with a bifurcate Killing horizon (with bifurcation surface $\Sigma$), and that the canonical mass and angular momentum of the solutions are well defined at infinity (see \cite{Wa1}). 
We shall apply this formulation to the spacetime we are considering, for which the entropy is given by
\begin{equation}
\label{eq:SW}
     S_{\mathrm{W}} =-2\pi\int_{\Sigma} \frac{\delta \mathcal{L}}{\delta R_{abcd}} 
                                                \epsilon_{ab}\epsilon_{cd} \sqrt{\mathbf{\mathrm{h}}} d\Omega_{3},
\end{equation}
where $\mathcal{L} = \mathcal{L}_{\mathrm{EGB}} + \mathcal{L}_{\text{\textrm{mat}}}$ is the sum of the Lagrangians corresponding to the EGB model and the matter source, respectively, 
$\sqrt{\mathrm{\mathbf{h}}} d\Omega_{3}$ is the volume element induced on $\Sigma$ by the metric of the spacetime, and $\epsilon_{ab}$ is the binormal tensor of the spatial section $\Sigma$ 
defined by the horizon given by $r=r_{h}$ and $t=\mathrm{cte}$, satisfying the normalization condition $\epsilon_{ab}\epsilon^{ab}=-2$. 
Giving two null (local) vector fields $\xi$ and $\nu$ normal to $\Sigma$ and satisfying that $g(\xi,\nu)=1$, the binormal tensor may be written as 
$\epsilon_{ab}=\xi_{a}\nu_{b} - \nu_{a}\xi_{b}$. 

We point out that in our case $\delta {\mathcal{L}}_{\text{\textrm{mat}}}/\delta R_{abcd}$=0, due to the fact that the string cloud matter is minimally coupled to gravity. 
Therefore, the matter term in \eqref{eq:SW} does not contribute to the entropy $S_{\mathrm{W}}$. 
It is not difficult to see that for the variation of the gravity action with respect to the Riemann tensor, regarding $R_{abcd}$ as formally independent to the metric $g_{ab}$, 
coincides with the entropy given in \eqref{eq:s}. 
A detailed derivation of this result can be found in \cite{ABD1} or \cite{GGS1}.

Moreover, using the Wald's entropy and the first law of thermodynamic  $dM_{\mathrm{W}}=T_{H} dS_{\mathrm{W}}$ we get, up to a constant $C_{\mathrm{0}}$, the mass of the black holes as 
\begin{equation}
\label{eq:MW}
     2 M_{\mathrm{W}}(r_{+}) = m(r_{+}) + C_{0}.
\end{equation}
Taking into account that we should obtain the Schwarzschild spacetime in the GR limit studied in Subsection \ref{subsec:g}, 
we conclude that the previous mass $2 M_{\mathrm{W}}$ is equal to the mass of the Schwarzschild black hole, which further implies that $C_{\mathrm{0}}$ vanishes. 
Hence, $M_{\mathrm{W}}$ coincides with the ADM mass $M$. 

The free energy $F = M - T_{\mathrm{H}} S$ of the black hole is given by 
\begin{equation}
\label{eq:fe}
    F = \frac{\frac{\Lambda}{3}r_{+}^{6} + 2 (1 + 6 \alpha \Lambda) r_{+}^{4} 
                - 8 \bar{a} r_{+}^{3} - 12 \alpha r_{+}^{2} 
                + 48 \alpha^{2}}
               {24 (r_{+}^{2} + 4 \alpha)}.     
\end{equation}
For large values of $r_{+}$, $F$ diverges to $\mathrm{sgn}(\Lambda)\infty$. 
On the other hand, for small values of $r_{+}$, $F$ goes to $\alpha$. 

In the limit $\bar{a} \rightarrow 0$, \eqref{eq:fe} gives the same expression found in \cite{Cai1}. 
For $\Lambda < 0$ we obtain that for small radius of event horizons the free energy is positive, 
so the black hole is \emph{globally unstable} (in the thermodynamical sense), 
whereas for large values of $r_{+}$ it is negative (\textit{i.e.} \emph{globally stable} black hole) (\textit{cf.}~\cite{AP1}). 
This is thus the Hawking-Page transition mentioned above. 
Besides, using the eq. \eqref{eq:fe}, we can explicitly obtain the critical temperature where the free energy is zero ($F(T_{\mathrm{c}})=0$). 
Then, the critical temperature in term of the parameters $\Lambda,a $ and the outer horizon $r_{+}$ is given by 
\begin{equation}
\label{eq:tcfe}
    T_{\mathrm{c}} = \frac{ -\Lambda r_{+}^{4} + 6 r_{+}^{2} +12 (\alpha -\bar{a} r_{+})}
               {24\pi (\frac{r_{+}^{3}}{3} + 4 \alpha r_{+})}.     
\end{equation}
The latter critical temperature essentially indicates the point where the HPt occurs. 
More precisely, for $T > T_{\mathrm{c}}$ the black hole solution is thermally globally preferred with respect to the reference background while for 
$T < T_{\mathrm{c}}$ the reference background solution is globally favored 
(for a detailed discussion  on this point with $\bar{a}=0$ and with Maxwell corrections, see \cite{AP1}).

\begin{figure}[H]
\begin{center}
\includegraphics[width=0.4\textwidth]{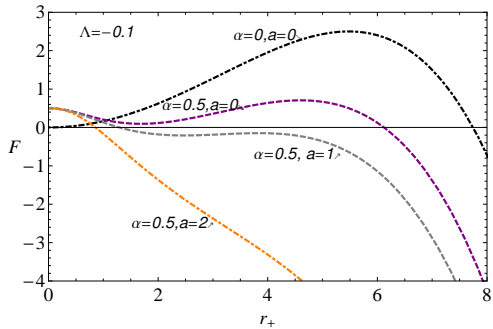}
\caption{The free energy as a function of $r_{+}$ for a fixed value of $\Lambda$ and several values of $a$ and $\alpha$. 
}
\label{fig:fig3}
\end{center}
\end{figure}

In Fig. \ref{fig:fig3} we indicate how the HPt is achieved by varying the value 
of the string cloud parameter $a$: 
if the free energy only admits one zero the system passes from an unstable phase to a stable regime. 
However, from a more careful analysis we see that the free energy may have three roots, so the system exhibits two stable phases: 
one for the small radius and the other for the large horizon radius. 
Besides, we obtain that the first stability zone is enlarged when increasing the parameter $a$ over the range 
$[0.5, 1.3]$. 

\begin{figure}[H]
\begin{center}
\includegraphics[width=0.4\textwidth]{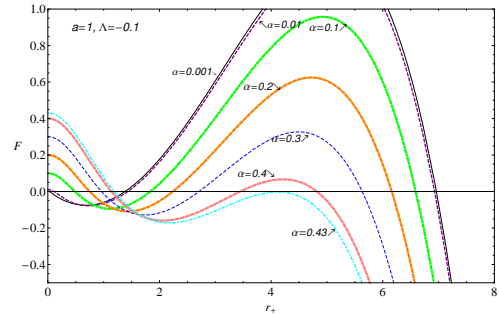}
\caption{The free energy $F$ as a function of $r_{+}$ for a fixed value of $a=1$ and $\Lambda=-0.1$.
}
\label{fig:fig4}
\end{center}
\end{figure}

In Fig. \ref{fig:fig4} we fix $\Lambda=-0.1$ and $a=1$. 
We examine $F(r_{+})$ when the Gauss-Bonnet parameter varies over the range $[0.001, 0.43]$. 
It turns out that the function $F(r_{+})$ has the same phase structure, U-S-U-S (U denotes an unstable zone and S a stable one), then our model seems not only to present the standard HPt but also to offer new phases for the global stability of the EGB gravity. 
Moreover, the first stability region corresponding to small horizon radii is amplified when increasing $\alpha$. 

Here we have only dealt with the case $\alpha \geq 0$ because our model is more related to string theory. 
However, some authors have considered the EGB model for $\alpha < 0$ and its implications in the thermodynamics of black hole solutions (see \cite{MKP1}). 

\subsection{\texorpdfstring{Local stability}{Local stability}}

Despite the fact that we have studied the condition for the global stability of the solution we are also going to examine the local thermodynamics structure by computing the specific heat. 
The reason is that even when a black hole configuration is found to be globally stable, it could also be locally unstable to some globally favored configuration (\textit{cf.}~\cite{AP1} and references therein).

The heat capacity $C = \partial M/\partial T$ turns out to be 
\begin{equation}
\label{eq:cv}
     C = \frac{\pi (r_{\mathrm{h}}^{2}+4 \alpha)^{2}|r_{\mathrm{h}} -\bar{a}-\frac{\Lambda}{3} r_{\mathrm{h}}^{3}|}
                    {(1-\Lambda r_{\mathrm{h}}^{2})(r_{\mathrm{h}}^{2}+4 \alpha) - 2 r_{\mathrm{h}}(r_{\mathrm{h}} -\bar{a}-\frac{\Lambda}{3} r_{\mathrm{h}}^{3})}.    
\end{equation}
By simplicity, we shall only analyze the sign of $C$ for $\Lambda = 0$. 
For $\bar{a} \rightarrow 0$ we recover the specific heat formula reported in \cite{Cai1}. 

\begin{figure}[H]
\begin{center}
\includegraphics [angle=0,width=0.4\textwidth] {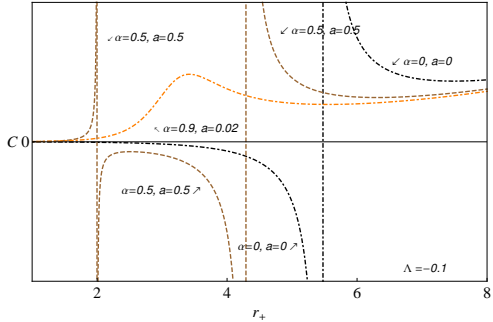}
\caption{Heap capacity $C$ as a function of $r_{+}$ for $\Lambda = -0.1$ with $(\alpha,a) = (0, 0)$, 
$(\alpha,a) = (0.5, 0.5)$ or $(\alpha,a) = (0.9, 0.02)$.}
\label{fig:cv}
\end{center}
\end{figure}

It is a well-known fact that the local thermodynamic stability of the system is related to the sign of the heat capacity. 
When $r_{\mathrm{h}} > \bar{a}$, we get that the black hole is \emph{locally stable} to thermal fluctuations (\textit{i.e.} $C> 0$) iff $m < \alpha$, 
whereas the heat capacity is negative (so the black hole is \emph{locally unstable}) iff $m > \alpha$. 
Notice that the point where the heat capacity vanishes is $\alpha = m$, corresponding to $r_{\mathrm{h}} = 2 \bar{a}$. 
Furthermore, in this case, the critical temperature is 
\[    T_{\mathrm{H}} = \frac{\bar{a}}{8 \pi (\bar{a}^{2} + \alpha)}.     \]
In the opposite case $r_{\mathrm{h}} < \bar{a}$, the thermodynamical behaviour is interchanged, \textit{i.e.} the solution is stable (resp. unstable) iff $m > \alpha$ 
(resp. $m < \alpha$). 
The jump of the heat capacity at the point $r_{\mathrm{h}} = 2 \bar{a}$ indicates a second order phase transition (see Fig. \ref{fig:cv}). 
Notice that in the GR limit ($\alpha \rightarrow 0$), the critical temperature remains finite. 

\subsection{\texorpdfstring{Entropy: Quantum corrections}{Entropy: Quantum corrections}}
\label{subsec:quantum}

In this subsection we shall analyze the quantum corrections to the classical entropy computed in \eqref{eq:s} for $\Lambda = 0$. 
Since we shall consider formal series on the Planck constant $\hbar$, we shall restore it in the previous formulas for the entropy. 

We shall denote $S_{\mathrm{BH}} = A/(4 \hbar)$ the entropy given by the Bekenstein-Hawking law 
and $S_{\mathrm{GB}}$ the entropy computed in the previous section (with the $\hbar$)
\[     S_{\mathrm{GB}} = \frac{A}{4 \hbar} \Big( 1 + \frac{12 \alpha}{r_{+}^{2}} \Big).     \]

In the semiclassical approximation for General Relativity, the black hole entropy satisfies the Bekenstein-Hawking area law, 
whereas when the full quantum effect is considered, this area law should be corrected. 
Further, the corrected entropy $S_{\mathrm{q}}$ takes the form
\begin{equation}
\label{eq:scorregida}
     S_{\mathrm{q}} = S + \theta \ln(S) + \dots,     
\end{equation}
where $S$ is the uncorrected semiclassical entropy of the black hole and $\theta$ is a dimensionless constant. 

Following the considerations on the tunneling method of \cite{BM1}, the corrected Hawking temperature $T_{\mathrm{q}}$ of a black hole is given by 
\[     T_{\mathrm{q}} = T_{\mathrm{H}} \Big( 1 + \sum_{i \geq 1} \gamma_{i} \hbar^{i} \Big)^{-1},     \]
where $T_{\mathrm{H}}$ is the standard semiclassical Hawking temperature of the black hole and for some (dimensional) constants $\gamma_{i}$. 

Whereas in the first three paragraphs of this section we have considered $\alpha$ and $\bar{a}$ as external parameters, 
we shall now allow them to vary within the model. 
This in turn implies that they should have a corresponding work-term in the expression of the first of thermodynamics, \textit{i.e.}
\[     dM = T_{\mathrm{H}} dS_{\textrm{GB}} + W_{\alpha} d\alpha + W_{\bar{a}} d{\bar{a}},     \]
where $W_{\alpha}$ and $W_{\bar{a}}$ are the work-terms of $\alpha$ and $\bar{a}$, respectively.  
At this point, one may argue why these entropy and temperature should coincide with the one we computed in the previous subsections. 
We leave the proof of this simple fact to the Appendix. 

Now, we see that, since the work-terms of $\alpha$ and $\bar{a}$ are not supposed to change by quantum effects, 
the semiclassical approximation leads to 
\begin{equation}
\label{eq:primerprincip}
    T_{\mathrm{H}} dS_{\textrm{GB}} = T_{\mathrm{q}} dS_{\mathrm{q}},     
\end{equation}
which means that in fact we need not compute the work-terms coming from the parameters $\alpha$ and $\bar{a}$. 
 
As noted in \cite{ZRL1}
, the constants $\gamma_{i}$ should be determined using dimensional considerations. 
In this case, since the following dimensions coincide $[\alpha] = [\bar{a}]^{2} = [r]^{2}$, we propose
\[     \gamma_{i} = \frac{\alpha_{i}}
                         {\Big(\sum\limits_{(\bar{j})} a_{\bar{j}}^{i} r_{+}^{j_{1}} \bar{a}^{j_{2}} \alpha^{\frac{3 - (j_{1} + j_{2})}{2}}\Big)^{i}},     \]
where $\alpha_{i}$ and $a_{\bar{j}}^{i}$ are dimensionless constants and the sum is indexed by $\bar{j} = (j_{1},j_{2}) \in \RR^{2}$ and 
of finite support. 
This is a generalization of the eq. (27) proposed in \cite{ZRL1}.

Hence, the corrected Hawking temperature has the form 
\begin{equation}
\label{eq:correcteds}
     T_{\mathrm{q}} = T_{\mathrm{H}} \left( 1 + \sum_{i \geq 1} \frac{\alpha_{i} \hbar^{i}}
                         {\Big(\sum\limits_{(\bar{j})} a_{\bar{j}}^{i} r_{+}^{j_{1}} \bar{a}^{j_{2}} \alpha^{\frac{3 - (j_{1} + j_{2})}{2}}\Big)^{i}}   \right)^{-1}.
\end{equation}

From equation \eqref{eq:primerprincip}, the differential of the corrected entropy is given by
\[     dS_{\mathrm{q}} = \frac{T_{\mathrm{H}}}{T_{\mathrm{q}}} \big((\pi r_{+}^{2} + 4 \pi \alpha) dr_{+} + 4 \pi r_{+} d\alpha \big),     \]
where we have used eq. \eqref{eq:s}. 
Since the entropy $S_{\mathrm{q}}$ is a state function, $dS_{\mathrm{q}}$ should be an exact differential, so 
\begin{align*}
   \frac{\partial}{\partial \alpha} \Big( \frac{T_{\mathrm{H}}}{T_{\mathrm{q}}}(\pi r_{+}^{2} + 4 \pi \alpha)\Big) 
   &= \frac{\partial}{\partial r_{+}} \Big( \frac{T_{\mathrm{H}}}{T_{\mathrm{q}}} 4 \pi r_{+} \Big), 
\\
   \frac{\partial}{\partial \bar{a}} \Big( \frac{T_{\mathrm{H}}}{T_{\mathrm{q}}}(\pi r_{+}^{2} + 4 \pi \alpha)\Big) &= 0,
   \frac{\partial}{\partial \bar{a}} \Big( \frac{T_{\mathrm{H}}}{T_{\mathrm{q}}} 4 \pi r_{+} \Big) = 0.
\end{align*}
From the last two equations we see that $T_{\mathrm{H}}/T_{\mathrm{q}}$ should not depend on $\bar{a}$, so 
\begin{equation}
\label{eq:correctedsnueva}
     T_{\mathrm{q}} = T_{\mathrm{H}} \left( 1 + \sum_{i \geq 1} \frac{\alpha_{i} \hbar^{i}}
                         {\Big(\sum\limits_{(j \in \RR)} a_{j}^{i} r_{+}^{j} \alpha^{\frac{3 - j}{2}}\Big)^{i}}   \right)^{-1}.
\end{equation}
The remaining equation implies that 
\[     \sum\limits_{(j \in \RR)}\Big( 6(j-1) a^{i}_{j} - \frac{5-j}{2} a^{i}_{j-2} \Big) r^{j}_{+} \alpha^{\frac{3-j}{2}} = 0.     \]
Hence, for all $j \in \RR$ we have that $12(j-1) a^{i}_{j} = (5-j) a^{i}_{j-2}$. 
Since the support of the sum was finite, we see that $a^{i}_{j} = 0$ for all $j < 1$ and $j > 5$. 
Moreover, by the form of the previous expression we conclude that $a^{i}_{j}$ vanishes if $j \notin 2\ZZ+1$ and if $j = 5$. 
The only non-trivial relation we obtain is $12 a^{i}_{3} = a^{i}_{1}$, which means that 
\[     \gamma_{i} = \frac{3\bar{\alpha}_{i}}{\pi(r_{+}^{3} + 12 r \alpha)^{i}},     \]
for some new dimensionless constant $\bar{\alpha}_{i}$ (\textit{cf.} \cite{ZRL1}, eq. (27) and eq. (32)).
So, 
\[     T_{\mathrm{q}} = T_{\mathrm{H}} \Big( 1 + \sum_{i \geq 1} \frac{\bar{\alpha}_{i}}{S_{\textrm{GB}}^{i}}  \Big)^{-1},     \]
which in turn implies that (using eq. \eqref{eq:primerprincip})
\[     S_{\mathrm{q}} = S_{\textrm{GB}} + \bar{\alpha}_{1} \log(S_{\textrm{GB}}) + \sum_{i \geq 2} \frac{\bar{\alpha}_{i}}{(1-i) S_{\textrm{GB}}^{i-1}} + \text{const}.     \]
This expression coincides with the one found in eq. (27) in \cite{ZRL1}, which seems reasonable since the entropy does not depend on the extra parameter $\bar{a}$. 
However, we would like to stress that we have adopted a more general assumption than \cite{ZRL1} for the expression of the constants 
in the semiclassical expansion of the Hawking temperature. 

\subsection{\texorpdfstring{Evaporation rate of the black hole}{Evaporation rate of the black hole}}
\label{subsec:eva}

We shall discuss the evaporation of the black hole with $\Lambda = 0$.
Taking into account the Stefan-Boltzmann law for the black body in $5$ dimensions we see that the evaporation rate is
\begin{equation}
\label{eq:er}
\frac{dm}{dt}\propto -T^{5}_{\mathrm{H}}r^{3}_{\mathrm{h}}.
\end{equation}
First, we see that we have a well-defined horizon radius if and only the mass parameter fulfills the requirement
\begin{equation}
\label{eq:hrdefinido}
  m \geq \mathrm{sup}.\{ 0 , \alpha - \bar{a}^{2}/2 \}.
\end{equation}  
This in turn implies that, if $\alpha - \bar{a}^{2}/2 > 0$, then the black hole cannot decay 
(this generalizes previous considerations in \cite{AFG1}). 

Otherwise, if $\alpha - \bar{a}^{2}/2 \leq 0$, the black hole could decay or not. 
More concretely, if $\alpha - \bar{a}^{2}/2 = 0$ then the evaporation time of the black hole is infinite, 
whereas if $\alpha - \bar{a}^{2}/2 < 0$ it evaporates in finite time for $\bar{a} > 0$ and in infinite time for $\bar{a}=0$. 
This can be proved as follows. 

Since $dm/dt =(r_{+}-\bar{a})dr_{+}/dt$, we may assume that, as $m$ varies from the initial mass parameter $m_{0}$ at time $t_{0}$ to $0$ at time $t_{e}$, $r_{+}$ goes from $r_{+,0}$ to $r_{+,e} = \bar{a}+\sqrt{\bar{a}^{2}-2 \alpha}$. 
Then, if we denote by $\Delta t = t_{e} - t_{0}$ the interval of time for that variation 
\[     \Delta t = \int_{r_{+,0}}^{r_{+,e}} \Big(\frac{dr_{+}}{dt}\Big)^{-1} dr_{+}
                \propto \int_{r_{+,e}}^{r_{+,0}} \frac{(r^{2}_{+}+4 \alpha)^{5}}{(r_{+}-\bar{a})^{4} r_{+}^{3}} dr_{+}.     \]    
If $\alpha - \bar{a}^{2}/2 = 0$, then one limit of integration is $r_{+,0} = \bar{a}$ and the integral diverges since the integrand goes as $(r_{+}-\bar{a})^{-4}$. 
Hence the evaporation time is infinite. 

On the other hand, if $\alpha - \bar{a}^{2}/2 < 0$, and assuming that $\bar{a} > 0$, we see that the integrand is  bounded and continuous within the integration interval, so the black holes evaporates in finite time. 
In the case $\bar{a} = 0$, we easily see that the integral diverges. 

\section{\texorpdfstring{Scalar waves propagating on string cloud background}{Scalar waves propagating on string cloud background}}
\label{sec:scalar}

In this section we shall consider the case of a scalar field $\Phi$ with mass $\mu$ on a $5$-dimensional spacetime 
with a string cloud background. 
We follow mainly \cite{IIV1}. 

The scalar field $\Phi$ satisfies the \emph{Klein-Gordon equation}
\[     \frac{1}{\sqrt{-\mathbf{g}}} \partial_{a}(\sqrt{-\mathbf{g}} g^{a b} \partial_{b}\Phi) - \mu^{2}\Phi = 0.     \]  
We shall assume that, in local coordinates, $\Phi(r,t,\phi,\theta,\xi) = \Psi(r,\theta,\xi) e^{-i (\omega t - l_{1} \phi)}$, where $l_{1} \in \ZZ$, 
and further that $\Psi(r,\theta,\xi) = R(r)Y(\theta)Z(\xi)$. 
Since the solution in the angular part is the same as in the classical case, we shall only analyze the radial part. 
In this case, there exist a constant $\bar{l}_{3} \in \NN_{0}$ such that the radial Klein-Gordon equation can be rewritten as 
\begin{align}
   \label{eq:r}
   \frac{\partial_{r}(r^{3}G\partial_{r}R)}{rR} + \Big(\frac{\omega^{2}}{G} - \mu^{2}\Big) r^{2} = \bar{l}_{3}^{2}.
\end{align}

Now we choose $\bar{R}(r) = r^{3/2} R(r)$ and the so called \emph{tortoise coordinate} $r_{*} = r_{*}(r)$ given by 
$ dr_{*}/dr = G(r)^{-1}$. 
We also set $R_{*}(r^{*}) = \bar{R}(r)$. 
Therefore, if we divide equation \eqref{eq:r} by $r^{-1/2}$, we get
\[     \frac{d^{2}R_{*}}{dr_{*}^{2}} + \Big(\omega^{2} - V(r)\Big) R_{*} = 0,     \] 
where
\[     V(r) = G(r) \Big(\frac{\bar{l}_{3}^{2}}{r^{2}} + \frac{3}{4 r^{2}} G + \frac{3 G'}{2 r} + \mu^{2}\Big).     \]
Using the expression for $G(r)$ given in \eqref{eq:solfinal} and the dimensionless variables $\hat{r} = r/\sqrt{m}$, $\hat{\Lambda} = \Lambda m$, 
$\hat{\alpha} = 2 \alpha/m$ and $\hat{a} = a/\sqrt{m}$, 
we obtain that $G(\hat{r}) = 1 + \frac{\hat{r}^{2}}{2 \hat{\alpha}} \Big( 1 - y(\hat{r}) \Big)$, 
where 
\[     y(\hat{r}) = \Big(1 + \frac{2 \hat{\alpha} \hat{\Lambda}}{3} 
                          + \frac{8 \hat{a} \hat{\alpha}}{3 \hat{r}^{3}} 
                          + \frac{8 \hat{\alpha}}{\hat{r}^{4}}\Big)^{1/2}.     \]
Let us now assume that $\Lambda = 0$. 
This implies that there are no terms of positive exponent in the expansion in $\hat{r}$ of the potential. 
Its third order approximation is 
\[     V(\hat{r}) \simeq \mu^{2} -\frac{2 \hat{a} \mu^{2}}{3 \hat{r}} + \Big( \bar{l}_{3}^{2} + \frac{3}{4} - 2 \mu^{2} \Big) \frac{1}{\hat{r}^{2}} 
                         - \frac{2 \hat{a} \bar{l}_{3}^{2}}{3 \hat{r}^{3}}.     \]

We see that the second order approximation coincides with the one in \cite{IIV1} (when $\hat{\Lambda} = \hat{\alpha} = \mu = 0$) 
and that the parameter $\hat{\alpha}$ does not provide any correction up to third order. 
We remark that the parameter $\hat{\alpha}$ appears in the fourth order term of the expansion.

\begin{figure}[H]
\begin{center}
\includegraphics [angle=0,width=0.4\textwidth] {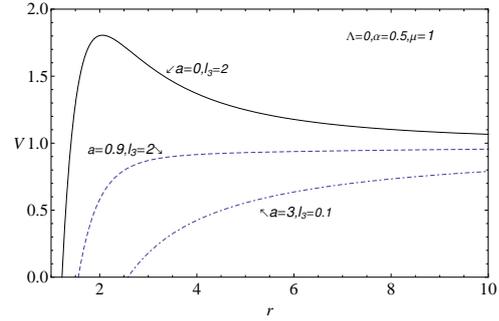}
\caption{Potential $V$ as a function of $r_{+}$ for a unit mass parameter $m = 1$ with fixed values of 
$\Lambda = 0$, $\alpha = 0.5$ and $\mu = 1$ and different values of $a$ and $l_{3}$.}
\label{fig:potencial2}
\end{center}
\end{figure}

Moreover, from the analysis of Fig. \ref{fig:potencial2}, we see that the potentials are real and positive outside the event horizon. 
Hence, following the argument given by S. Chandrasekhar in \cite{Cha1}, the black hole is stable under scalar perturbations. 
Besides, when a maximum of $V(\hat{r})$ exists, it seems to decrease as $\hat{a}$ increases. 

\section{\texorpdfstring{Conclusions}{Conclusions}}

To summarize, we have found black holes in Einstein-Gauss-Bonnet gravity for a static and spherically symmetric 5-dimensional spacetime with an energy momentum given by a cloud of strings. 
We characterized the solution, calculating the possible horizons, which could be at most two. 
Besides, we confirmed the singular structure of the spacetime at the origin by computing some quadratic invariants. 

We performed a detailed analysis of the more relevant thermodynamical aspects, focusing mainly on the global and local stability of the system. 
In particular, the entropy of the black hole was computed using the Wald prescription. 
We also found that the HPt can be realized within this context. 
Interestingly, the region for small stable black holes are considerably enlarged when choosing small values of 
$(\alpha, a)$. 
This has been already observed in the literature for other sources (\textit{cf.}~\cite{AP1}, \cite{Cai1}, \cite{CN1}, \cite{CNO1}). 

Regarding the local stability of the black holes, we obtained that the heat capacity can be negative or positive definite, depending on the relation between $m$ and $\alpha$, which further tells us whether the black hole is stable or not. 
In particular, the point where the heat capacity is zero corresponds to the case $r_{\mathrm{h}} = 2 \bar{a}$ and non-zero Hawking temperature. 
Notice that this critical temperature is zero for a vacuum solution $\bar{a} = 0$ (\textit{cf.}~\cite{BD1}, \cite{Cai1}). 

Later on, we examined the quantum correction to the Hawking temperature and to the entropy, 
following the prescriptions of tunneling method used in \cite{BM1}, \cite{ZRL1}. 
We have provided a proof of a lemma which tells us that the expression for the GB entropy found 
in the case that $\alpha$ and $a$ are treated as external parameters holds when these parameters are allowed to vary. 

After that, we estimated the evaporation rate for the black hole
, obtaining that for $2 \alpha > \bar{a}^2$ the black hole cannot decay for its mass parameter is bounded from below. However, for $2\alpha < \bar{a}^2$ the black hole evaporates in finite time if $\bar{a} > 0$, 
whereas the evaporation time is infinite if $\bar{a}=0$. 
For the critical case corresponding to $2\alpha = \bar{a}^2$ we showed that the evaporation time is infinite.

Finally, we studied the propagation of scalar waves using this metric as a background. 
We basically found that when the potential $V(r)$ has a maximum, the black holes are stable. 

\section*{Acknowledgments}

We are grateful to the referee for his careful reading of the manuscript.
This work has been partially supported by the projects UBACyT X212 and PICT-2007-02182. 
The first author is a research fellow of the Alexander von Humboldt Foundation. 
The second author is a CONICET fellow (Argentina).

\section*{Appendix}
\label{sec:ap}

In this appendix we shall prove the elementary result stated in Lemma \ref{lem:lema}. 
We may rephrase it in more general terms (\textit{e.g.} for Banach manifolds) but we prefer to stay concrete. 
We first need to state some notation. 

We fix $p \geq 1$. 
Let us suppose that two $C^{p}$ functions $f : U \subset \RR^{n+m} \rightarrow \RR$ and $g : U \subset \RR^{n+m} \rightarrow \RR^{n^{2}} = M_{n}(\RR)$ 
defined on an open set $U$ are given, such that $\det(g)$ is non-vanishing. 
We shall denote an element of $\RR^{n+m}$ as $(\bar{x},\bar{y})$, with $\bar{x} \in \RR^{n}$ and $y \in \RR^{m}$. 

We are interested in a solution to the following problem: 
find $C^{p}$-morphisms $\bar{f} : V \subset \RR^{n+m} \rightarrow \RR$ and $h : U \subset \RR^{n+m} \rightarrow \RR^{n}$ defined on an open set $V$ such that 
the map $H : U \rightarrow V$ given by $(\bar{x},\bar{y}) \mapsto (h(\bar{x},\bar{y}),\bar{y})$ is $C^{p}$-isomorphism and the following identities 
hold in $U$
\begin{equation}
\label{eq:1}
\begin{cases}
\frac{\partial \bar{f}}{\partial \bar{x}} (h(\bar{x},\bar{y}),\bar{y}) = g(\bar{x},\bar{y}),
\\
\bar{f}(h(\bar{x},\bar{y}),\bar{y}) = f(\bar{x},\bar{y}).
\end{cases}
\end{equation}
A problem as before will be called \emph{a change of domain problem}. 
Notice that $(\bar{f},h)$ are not independent since $\bar{f} = f \circ H^{-1}$. 
We also see that 
\begin{equation}
\label{eq:2}
     \frac{\partial h}{\partial \bar{x}} (\bar{x},\bar{y}) = g(\bar{x},\bar{y})^{-1} \frac{\partial f}{\partial \bar{x}} (\bar{x},\bar{y}).     
\end{equation}
Furthermore, the fact that $H$ is a $C^{p}$-isomorphism implies that the previous expression is an invertible matrix. 

\begin{lema}
\label{lem:lema}
Let us suppose that two $C^{p}$ functions $f : U \subset \RR^{n+m} \rightarrow \RR$ and 
$g : U \subset \RR^{n+m} \rightarrow \RR^{n^{2}} = M_{n}(\RR)$ defined on an open set $U$, are given, such that 
$\det(\partial f/\partial \bar{x} (\bar{x}_{0},\bar{y}_{0}))$ and $\det(g(\bar{x}_{0},\bar{y}_{0}))$ do not vanish 
for a fixed point $(\bar{x}_{0},\bar{y}_{0}) \in U$. 
Then a local solution to the change of domain problem in a neighbourhood of $(\bar{x}_{0},\bar{y}_{0})$ exists. 
Moreover, any two (global) solutions to the change of domain problem $(\bar{f}_{1},h_{1})$ and $(\bar{f}_{2},h_{2})$ satisfy 
that $h_{1}-h_{2}$ is a function of $\bar{y} \in \RR^{m}$. 
\end{lema}
\noindent\textbf{Proof.}
Using the Existence and Uniqueness Theorem for ODE's in eq. \eqref{eq:2}, we see that there exists a locally defined function $h$ as needed. 
The Inverse Function Theorem tells that the map $H$ given by $(\bar{x},\bar{y}) \mapsto (h(\bar{x},\bar{y}),\bar{y})$ is a local $C^{p}$-isomorphism 
at $(\bar{x}_{0},\bar{y}_{0})$. 
We restrict the domain in order to assure this last property. 
Now we define $\bar{f} = f \circ H^{-1}$. 

We need to check that the first equation in \eqref{eq:1} holds. 
However, from the chain rule and the Inverse Function Theorem, it is easy to see that the first equation in \eqref{eq:1} is equivalent 
to \eqref{eq:2}, and so the local existence follows.

Also, we see that if a solution to the previous problem exists, it is not unique. 
This is due to the following simple fact: if $k : \RR^{m} \rightarrow \RR^{n}$ is a $C^{p}$-morphism, we may define 
$\tilde{h}(\bar{x},\bar{y}) = h(\bar{x},\bar{y}) + k(\bar{y})$. 
It is trivially verified that the map $\tilde{H}$ given by $(\bar{x},\bar{y}) \mapsto (\tilde{h}(\bar{x},\bar{y}),\bar{y})$ is a $C^{p}$-isomorphism, 
being the composition of $H$ and the $C^{p}$-isomorphism $(\bar{x},\bar{y}) \mapsto (\bar{x}+k(\bar{y}),\bar{y})$ 
(with inverse $(\bar{x},\bar{y}) \mapsto (\bar{x}-k(\bar{y}),\bar{y})$). 
We may now define $\tilde{f}(\bar{x},\bar{y}) = \bar{f}(\bar{x}-k(\bar{y}),\bar{y})$. 
It is trivially verified that $(\tilde{f}, \tilde{h})$ is also a solution to the change of domain problem. 
Moreover, given any two solutions to the change of domain problem, they are related as before, since equation \eqref{eq:1} implies that 
\[     \frac{\partial h}{\partial \bar{x}} (\bar{x},\bar{y}) 
       = (g(\bar{x},\bar{y}))^{-1} \frac{\partial f}{\partial \bar{x}} (\bar{x},\bar{y}).     \]
The lemma is thus proved. 
\qed

We may apply the previous lemma to the case $(n,m) = (1,2)$, $\bar{x} = r_{+}$, $\bar{y} = (\alpha, \bar{a})$, $f = M(r_{+},\alpha,\bar{a})$, and 
$g = T_{\mathrm{H}}(r_{+},\alpha,\bar{a})$. 
In this case, it it easy to see that the function $\bar{f} = S$ given by eq. \eqref{eq:s} (and a certain $h$) provides a solution to the change of 
domain problem. 
Furthermore, by the previous lemma and the physical assumption that the entropy vanishes for $r_{+} = 0$ we conclude that the expression for the entropy 
given by \eqref{eq:s} is unique and it also holds when allowing the parameters $\alpha$ and $\bar{a}$ to vary.

\end{document}